\documentclass[11pt]{article}

\pdfoutput=1

\usepackage[latin1]{inputenc}
\usepackage{amscd,amsmath,amssymb,amsfonts,latexsym,amsthm,mathrsfs,bbm}
\usepackage{graphicx,color}
\usepackage{enumerate}
\usepackage{natbib}
\bibpunct{(}{)}{,}{a}{,}{,}
\usepackage{verbatim}
\usepackage{fullpage}

\newcommand{\bx}{{\mathbf{x}}}
\newcommand{\tFo}{{{\,}_{2} F_{1}}}

\newcommand{\by}{\mathbf{y}}
\newcommand{\bz}{{\mathbf{z}}}
\newcommand{\bX}{\mathbf{X}}
\newcommand{\bP}{\mathbf{P}}
\newcommand{\BE}{\mathbb{E}}
\newcommand{\E}{\mathbb{E}}
\newcommand{\bbeta}{{\boldsymbol{\beta}}}

\newcommand{\bgamma}{{\boldsymbol{\gamma}}}
\newcommand{\bGamma}{{\boldsymbol{\Gamma}}}

\newcommand{\ds}{\displaystyle}

\newcommand{\vs}{\vspace{0.5cm}}

\begin{document}

\title{Regularization in regression: comparing Bayesian and frequentist methods in a poorly informative situation
\footnote{This paper is part of Mohammed EL Anbari's PhD thesis. This work has been partly supported 
by the Agence Nationale de la Recherche (ANR, 212, rue de Bercy 75012 Paris) through the 2009-2012 project
ANR-09-BLAN-01 {\sf EMILE} for the last two authors, and by Institut Universitaire de France for the last author.
Jean-Michel Marin and Christian P.~Robert are grateful to the participants to the BIRS {\sf 07w5079}
meeting on ``Bioinformatics, Genetics and Stochastic Computation: Bridging the Gap''
for their helpful comments. Discussions in 2007 in Banff with Sylvia Richardson and in
Roma with Jim Berger and Paul Speckman are also gratefully acknowledged. Given that Arnold Zellner sadly passed away
last August, we would like to dedicate this paper to the memory of this leading Bayesian thinker who influenced so much
the field and will continue to do so much longer.}}

\author{{\sc Gilles Celeux} \\
{\em Project \textsc{select}, INRIA Saclay, Universit\'e Paris Sud, Orsay, France} \\
{\sc Mohammed EL Anbari} \\
{\em Universit\'e Caddi Ayyad, Marrakech, Maroc} \\
{\sc Jean-Michel Marin} \\
{\em Institut de Mathématiques et Modélisation de Montpellier,} \\
{\em Universit\'e de Montpellier 2, France} \\
\&\ {\sc Christian P. Robert} \\
{\em Universit\'e Paris-Dauphine, CEREMADE,} \\
{\em Institut Universitaire de France \& CREST, France}
}

\maketitle

\begin{abstract}
Using a collection of simulated an real benchmarks, we compare Bayesian and frequentist
regularization approaches under a low informative constraint when the number of variables
is almost equal to the number of observations on simulated and real datasets.
This comparison includes  new global noninformative approaches for Bayesian
variable selection  built on Zellner's g-priors that are similar to
\cite{Liang:Molina:Clyde:Berger:2008}. The interest of those calibration-free proposals is discussed.
The numerical experiments we present highlight the appeal of Bayesian regularization methods, when compared
with non-Bayesian alternatives. They dominate frequentist methods in the sense that they provide smaller prediction 
errors while selecting the most relevant variables in a parsimonious way.

\vs \noindent
\end{abstract}

\newpage

\normalsize
\section{Introduction}
Given a response variable, $y$ and a collection of $p$ associated potential predictor variables
$x_1,\ldots,x_p$, the classical linear regression model imposes a linear dependence on the conditional
expectation \citep{Rao:1973}
$$
\BE[y|x_1,\ldots,x_p]=\beta_0+\beta_1 x_1+\ldots \beta_P x_p\,.
$$
A fundamental inferential direction for those models relates to the variable selection
problem, namely that only variables of relevance should be kept within the regression while
the others should be removed. While we cannot discuss at length the potential applications
of this perspective, variable selection is particularly relevant when the number $p$ of regressors is
larger than the number $n$ of observations (as in microarray and other genetic data analyzes).

To deal with poorly or ill-posed regression problems, many regularization methods
have been proposed, like ridge regression \citep{Hoerl:Kennard:1970} and Lasso \citep{Tibshirani:1996}. 
Recently the interest for frequentist regularization methods has increased and this has
produced a flury of methods (see, among others, \citealp{Candes:Tao:2007,Zou:Hastie:2005,Zou:2006,Yuan:Lin:2007}).

However, a natural approach for regularization is to follow the Bayesian paradigm as demonstrated recently by
the Bayesian Lasso of \cite{Park:Casella:2008}. The amount of literature on Bayesian variable selection is 
quite enormous (a small subset of which is, for instance, \citealp{Mitchell:Beauchamp:1988,George:McCulloch:1993,
Chipman:1996,Smith:Kohn:1996,George:McCulloch:1997,Dupuis:Robert:2003,
Brown:Vannucci:1998,Philips:Guttman:1998,George:2000,Kohn:Smith:Chan:2001,
Nott:Green:2004,Schneider:Corcoran:2004,
Casella:Moreno:2006,Cui:George:2008,Liang:Molina:Clyde:Berger:2008,Bottolo:Richardson:2010}). The number of approaches
and scenarii that have been advanced to undertake the selection of the most relevant variables given a set of
observations is quite large, presumably due to the vague decisional setting induced by
the question {\em Which variables do matter?} Such a variety of resolutions signals a
lack of agreement between the actors in the field.
Most of the solutions, including \cite{Liang:Molina:Clyde:Berger:2008} and \cite{Bottolo:Richardson:2010}, focus on the use of the $g$-prior,
introduced by \cite{Zellner:1986}.
While this prior has a long history and while it reduces the prior input to
a single integer, $g$,  the influence of this remaining prior
factor is long-lasting and  large values of $g$ are no guarantee of negligible
effects, in connection with the Bartlett or Lindley--Jeffreys paradoxes \citep{Bartlett:1957,
Lindley:1957,Robert:1993}, as illustrated for instance
in \cite{Celeux:Marin:Robert:2006} or \cite{Marin:Robert:2007}. In order to alleviate this
influence,  some empirical Bayes [\cite{Cui:George:2008}] and hierarchical Bayes
[\cite{Zellner:Siow:1980}, \cite{Celeux:Marin:Robert:2006}, \cite{Marin:Robert:2007}, \cite{Liang:Molina:Clyde:Berger:2008} 
and \cite{Bottolo:Richardson:2010}] solutions have been proposed.
In this paper, we pay special attention to two calibration-free hierarchical Zellner $g$-priors.
The first  one is the Jeffreys prior which is not location invariant. A second one avoids this problem
by only considering models with at least one variable in the model. 
%a  natural Bayesian path is 
%proceed to the construction of a noninformative
%alternative that eliminates the impact of the factor $g$,
%based on the Jeffreys prior associated with a hierarchical model.
%We thus claim to have achieved here a calibration-free Bayesian approach to the variable selection problem.
%Thus, a calibration-free Bayesian approach to the variable selection problem is proposed and assessed. 

The purpose of our paper is to compare the frequentist and the Bayesian
points of views in regularization when $n$ remains (slightly) greater than $p$,
% As a matter of fact, the use of $g$-prior implies that the sample size $n$ has to be greater than $p$.
we limit our attention to full rank models. This comparison is considered from both the predictive and
the explicative point of views. The outcome of this study is that Bayesian methods are quite similar
while dominating their frequentist counterpart. 

\newpage

The plan of the paper is as follows: we recall the details of Zellner's (\citeyear{Zellner:1986})
original $g$-prior in Section \ref{sec:gee}, and discuss therein the potential choices of $g$.
We present hierarchical noninformative alternatives in Section \ref{sec:nonigee}. Section \ref{sec:simu}
compares the results of Bayesian and frequentist methods on simulated and real datasets. Section \ref{sec:conc}
concludes the paper.

\section{Zellner's $g$-priors}\label{sec:gee}

Following standard notations, we introduce a variable $\bgamma\in\bGamma=\{0,1\}^{\otimes p}$
that indicates which variables are active in the regression, excluding the constant vector corresponding to
the intercept that is assumed to be always present in the linear regression model.

We observe $\by,\bx_1,\ldots,\bx_p \in \mathbb{R}^n$, the model $\mathcal{M}_{\bgamma}$ is defined as the
conditional distribution
\begin{equation}
\by|\bX,\bgamma,\bbeta^{\bgamma},\sigma^2\sim\mathcal{N}_n\left(\bX^\bgamma\boldsymbol{\boldsymbol{\beta}}^\bgamma,\sigma^2I_n\right)\,,
\end{equation}
where
\begin{itemize}
\renewcommand{\labelitemi}{$\blacktriangleright$}
\item $p_\bgamma=\sum_{i=1}^p \gamma_i$,
\item $\bX^{\bgamma}$ is the $(n,p_\bgamma+1)$ matrix which columns are made of
the vector $\mathbf{1}_n$ and of the variables $\bx_i$ for which $\gamma_i=1$,
\item $\bbeta^{\bgamma}\in\mathbb{R}^{p_{\bgamma}+1}$ and $\sigma^2\in\mathbb{R}^*_+$ are unknown parameters.
\end{itemize}
The same symbol for the parameter $\sigma^2$ is used across all models. For model $\mathcal{M}_{\bgamma}$, Zellner's $g$-prior is given by
$$
\boldsymbol{\beta}^\bgamma|\bX,\bgamma,\sigma^2\sim\mathcal{N}_{p_\bgamma+1}(\tilde{\boldsymbol{\beta}}^\bgamma,g_\bgamma\sigma^2((\bX^\bgamma)'\bX^\bgamma)^{-1})\,,
$$
$$
\pi(\sigma^2|\bX,\bgamma)\propto \sigma^{-2}\,.
$$
The experimenter chooses the prior expectation $\tilde{\boldsymbol{\beta}}^\bgamma$ and $g_\bgamma$.
For such a prior, we obtain the classical average between prior and observed regressors,
$$
\E(\boldsymbol{\beta}^\bgamma|\bX,\bgamma,\by)=\frac{g_\bgamma\hat{\boldsymbol{\beta}}^\bgamma+\tilde{\boldsymbol{\beta}}^\bgamma}{g_\bgamma+1}\,.
$$
This prior is traditionally called Zellner's $g$-prior in the Bayesian folklore because of the
use of the constant $g_{\bgamma}$ by \cite{Zellner:1986} in front of Fisher's information matrix
$((\bX^\bgamma)'\bX^\bgamma)^{-1}$.
Its appeal is that, by using the information matrix as a global scale,
\begin{itemize}
\renewcommand{\labelitemi}{$\blacktriangleright$}
\item it avoids the specification of a whole prior covariance matrix, which would be a tremendous task;
\item it allows for a specification of the constant $g_{\bgamma}$ in terms of observational units,
or virtual prior pseudo-observations in the sense of \cite{deFinetti:1972}.
\end{itemize}
However, fundamental feature of the $g$-prior is that this prior is improper, due to the use of an infinite mass
on $\sigma^2$. From a theoretical point of view, this should jeopardize the use of posterior model probabilities
since these probabilities are not uniquely scaled under improper priors, because there is no way of eliminating the
residual constant factor in those priors \citep{DeGroot:1973,Kass:Raftery:1995,Robert:2001}. However, under the assumption that
$\sigma^2$ is a parameter that has a meaning common to all models $\mathcal{M}_{\bgamma}$,
\cite{Berger:Pericchi:Varshavsky:1998} develop a framework that allows to work with a single
improper prior that is common to all models (see also \citealp{Marin:Robert:2007}).
A fundamental appeal of Zellner's $g$-prior in model comparison and in particular in variable selection is its simplicity,
since it reduces the prior input to the sole specification of a scale parameter $g$.

\vs At this stage, we need to point out that an alternative $g$-prior is often used
\citep{Berger:Pericchi:Varshavsky:1998,Fernandez:Ley:Steel:2001,Liang:Molina:Clyde:Berger:2008,Bottolo:Richardson:2010}, by
singling out the intercept parameter in the linear regression. By first assuming a centering of the covariates,
i.e.~$\mathbf{1}_n'\bx_i=0$ for all $i$'s, the intercept $\alpha$ is given a
flat prior while the other parameters of $\bbeta^\bgamma$ are associated with a corresponding $g$-prior.
Thus, this is an alternative to model $\mathcal{M}_{\bgamma}$, which we denote by model $\mathcal{M}_{\bgamma}^{\text{inv}}$
to stress the distinctions between both representations and which is such that
\begin{equation}
\by|\bX,\bgamma,\alpha,\boldsymbol{\beta}_{\text{inv}}^\bgamma,\sigma^2\sim\mathcal{N}_n\left(\alpha\mathbf{1}_n+\bX_{\text{inv}}^\bgamma\boldsymbol{\beta}_{\text{inv}}^\bgamma,\sigma^2I_n\right)\,,
\end{equation}
where
\begin{itemize}
\renewcommand{\labelitemi}{$\blacktriangleright$}
\item $\bX_{\text{inv}}^\bgamma$ the $(n,p_\bgamma)$ matrix which columns are made of the variables $\bx_i$ for which $\gamma_i=1$,
\item $\alpha\in\mathbb{R}$, $\boldsymbol{\beta}_{\text{inv}}^\bgamma\in\mathbb{R}^{p_\bgamma}$ and $\sigma^2\in\mathbb{R}^*_+$ are unknown parameters.
\end{itemize}
The parameters $\sigma^2$ and $\alpha$ are denoted the same way across all models and rely on the same prior.
Namely, for model $\mathcal{M}_{\bgamma}^{\text{inv}}$, the corresponding Zellner's $g$-prior is given by
$$
\boldsymbol{\beta}_{\text{inv}}^\bgamma|\bX,\bgamma,\sigma^2\sim\mathcal{N}_{p_\bgamma}(\tilde{\boldsymbol{\beta}}_{\text{inv}}^\bgamma,g_\bgamma
\sigma^2((\bX_{\text{inv}}^\bgamma)'\bX_{\text{inv}}^\bgamma)^{-1})\,,
$$
$$
\pi(\alpha,\sigma^2|\bX,\bgamma)\propto \sigma^{-2}\,.
$$
In that case, we obtain
$$
\E(\boldsymbol{\beta}_{\text{inv}}^\bgamma|\bX,\bgamma,\by)=\frac{g_\bgamma\hat{\boldsymbol{\beta}}_{\text{inv}}^\bgamma+\tilde{\boldsymbol{\beta}}_{\text{inv}}^\bgamma}{g_\bgamma+1}\,,
$$
and
$$
\E(\alpha|\bX,\bgamma,\by)=\bar\by=\frac{1}{n}\sum_{i=1}^n y_i\,.
$$

For models  $\mathcal{M}_{\bgamma}$ and $\mathcal{M}_{\bgamma}^{\text{inv}}$, in a noninformative setting, we can for instance choose $\tilde\bbeta^\bgamma=0_{p_{\bgamma}+1}$ or 
$\tilde\bbeta_{\text{inv}}^\bgamma=0_{p_{\bgamma}}$ and
$g_{\bgamma}$ large. However, as pointed out in \citeauthor{Marin:Robert:2007} (2007, Chapter 3)
among others, there is a lasting influence of $g_{\bgamma}$ over the resulting inference and it is impossible
to ``let $g_{\bgamma}$ go to infinity'' to  eliminate this influence, because of the Bartlett and Lindley-Jeffreys
\citep{Bartlett:1957,Lindley:1957,Robert:1993}
paradoxes that an infinite value of $g_{\bgamma}$ ends up selecting the null model, regardless of the information
brought by the data. For this reason, data-dependent versions of $g_{\bgamma}$ have been proposed
with various degrees of justification:
\begin{itemize}
\renewcommand{\labelitemi}{$\blacktriangleright$}
\item \cite{Kass:Wasserman:1995} use $g_{\bgamma}=n$ so that the amount of information about the
parameters contained in the prior equals the amount of information brought by
one observation. As shown by \cite{Foster:George:1994}, for $n$ large enough this perspective is very
close to using the Schwarz \citep{Kass:Wasserman:1995} or BIC criterion in that the log-posterior
corresponding to $g=n$ is equal to the penalized log-likelihood of this criterion.
\item \cite{Foster:George:1994} and \cite{George:Foster:2000} propose $g_{\bgamma}=p_{\bgamma}^2$, in connection
with the Risk Inflation Criterion (RIC) that penalizes the regression sum of squares.
\item \cite{Fernandez:Ley:Steel:2001} gather both perspectives in
$g_{\bgamma}=\max(n,p_{\bgamma}^2)$ as a conservative bridge between BIC and RIC, a choice that they christened ``benchmark
prior''.
\item \cite{George:Foster:2000} and \cite{Cui:George:2008} resort to empirical Bayes techniques.
\end{itemize}

These solutions, while commendable since based on asymptotic properties (see in particular
\citealp{Fernandez:Ley:Steel:2001} for consistency results), are nonetheless
unsatisfactory in that they depend on the sample size and involve a degree of arbitrariness.
%while a complete Bayesian solution is readily available, as demonstrated below.

\section{Mixtures of $g$-priors}\label{sec:nonigee}

The most natural Bayesian approach to solving the uncertainty on 
the parameter $g_{\bgamma}=g$ is  to put a hyperprior on this parameter:
\begin{itemize}
\renewcommand{\labelitemi}{$\blacktriangleright$}
\item This was implicitely proposed by \cite{Zellner:Siow:1980} since those authors introduced Cauchy priors
on the $\bbeta^\bgamma$'s since this corresponds to a $g$-prior augmented by a Gamma $\mathcal{G}a(1/2,n/2)$ prior on $g^{-1}$.
\item For model $\mathcal{M}_{\bgamma}^{\text{inv}}$, \cite{Liang:Molina:Clyde:Berger:2008}, \cite{Cui:George:2008}
and \cite{Bottolo:Richardson:2010} use
$$
\boldsymbol{\beta}_{\text{inv}}^\bgamma|\bX,\bgamma,\sigma^2\sim\mathcal{N}_{p_\bgamma}(0_{p_{\bgamma}},g
\sigma^2((\bX_{\text{inv}}^\bgamma)'\bX_{\text{inv}}^\bgamma)^{-1})
$$
and an hyperprior of the form
$$
\pi(\alpha,\sigma^2,g|\bX,\bgamma)\propto (1+g)^{-a/2}\sigma^{-2}\,,
$$
with $a>2$ . This constraint on $a$ is due to the fact that the
hyperprior must be proper, in connection with the separate processing of the intercept $\alpha$ and the use of a Lebesgue measure as a
prior on $\alpha$. We note that $a$ needs to be specified, $a=3$ and $a=4$ being the solutions favored by
\cite{Liang:Molina:Clyde:Berger:2008}.
\item For model $\mathcal{M}_\bgamma$, \cite{Celeux:Marin:Robert:2006} and \cite{Marin:Robert:2007} used
$$
\boldsymbol{\beta}^\bgamma|\bX,\bgamma,\sigma^2\sim\mathcal{N}_{p_\bgamma+1}(0_{p_{\bgamma}+1},g\sigma^2((\bX^\bgamma)'\bX^\bgamma)^{-1})
$$
and a hyperprior of the form
$$
\pi(\sigma^2,g|\bX)\propto \sigma^{-2}g^{-1}\mathbb{I}_{\mathbb{N}^*}(g)\,.
$$
The choice of the integer support is mostly computational, while the Jeffreys-like $1/g$ shape
is not justified, but the authors claim that it is appropriate for a scale parameter.
\end{itemize}

For model $\mathcal{M}_\bgamma$ a more convincing modelling is possible since  the Jeffreys prior is available.
Indeed, if
$$
\boldsymbol{\beta}^\bgamma|\bX,\bgamma,\sigma^2\sim\mathcal{N}_{p_\bgamma+1}(0_{p_{\bgamma}+1},g\sigma^2((\bX^\bgamma)'\bX^\bgamma)^{-1})\,,
$$
then
$$
\by|\bX,\bgamma,g,\sigma^2\sim\mathcal{N}_{p_\bgamma+1}\left(0_n,
\sigma^2\left[\mathbf{I}_n-\frac{g}{g+1}\bP_\bgamma\right]^{-1}\right)\,,
$$
where $\bP_\bgamma$ is the orthogonal projector on the linear subspace spanned by the columns of $\bX^\bgamma$.
Since, the Fisher information matrix is
$$
\mathfrak{I}(\sigma^2,g)=\left(\frac{1}{2}\right)\left[\begin{array}{cc}
 n\big/\sigma^4                   & (p_\bgamma+1)\big/(\sigma^2(g+1)) \\
 (p_\bgamma+1)\big/(\sigma^2(g+1)) & (p_\bgamma+1)\big/(g+1)^2
\end{array}\right],
$$
the corresponding Jeffreys prior on $(\sigma^2,g)$ is
$$
\pi(\sigma^2,g|\bX)\propto \sigma^{-2}(g+1)^{-1}\,.
$$
Note that, for model $\mathcal{M}_{\bgamma}^{\text{inv}}$, \cite{Liang:Molina:Clyde:Berger:2008}
discuss the choice of $a=2$ and then $\pi(\alpha,\sigma^2,g|\bX,\bgamma)\propto (1+g)^{-1}\sigma^{-2}$
as leading to the reference prior and Jeffreys prior, presumably also under the marginal model after
integrating out $\boldsymbol{\beta}^\bgamma$, although details are not given.

For such a prior modelling,  there exists a closed-form representation for posterior quantities in that
$$
\pi(\bgamma,g|\bX,\by)\propto (g+1)^{n/2-(p_\bgamma+1)/2-1}(1+g(1-\by'\bP_\bgamma\by/\by'\by))^{-n/2}
$$
and
\begin{equation}
\pi(\bgamma|\bX,\by)\propto \frac{\tFo(n/2,1;(p_\bgamma+3)/2;\by'\bP_\bgamma
\by\big/\by'\by)}{p_\bgamma+1}\,,
\label{eq:mpost1}
\end{equation}
where $\tFo$ is the Gaussian hypergeometric function \citep{Butler:Wood:2002}.
We can thus proceed to undertake Bayesian variable selection without resorting at all to numerical methods \citep{Marin:Robert:2007}.
Moreover, the shrinkage factor due to the Bayesian modelling can also be expressed in closed form as
\begin{eqnarray*}
\E(g/(g+1)|\bX,\bgamma,\by)
&=& \frac{\ds \int_0^\infty g(g+1)^{n/2-(p_\bgamma+1)/2-2}(1+g(1-\by'\bP_\bgamma\by/\by'\by))^{-n/2}\text{d}g}
{\ds \int_0^\infty (g+1)^{n/2-(p_\bgamma+1)/2-1}(1+g(1-\by'\bP_\bgamma\by/\by'\by))^{-n/2}\text{d}g} \\
&=& \frac{2\tFo(n/2,2;(p_\bgamma+3)/2+1;\by'\bP_\bgamma
\by\big/\by'\by)}{(p_\bgamma+3)\tFo (n/2,1;(p_\bgamma+3)/2;\by'\bP_\bgamma \by\big/\by'\by)}.
\end{eqnarray*}
This obviously leads to straightforward representations for Bayes estimates.
%\begin{eqnarray*}
%\hat\by_\text{new}^\bgamma
%&=&\E\left[\by_\text{new}|\bX_\text{new},\bX,\bgamma,\by\right]\\
%&=&2\,\frac{\tFo (n/2,2;(p_\bgamma+3)/2+1;\by'\bP_\bgamma \by\big/\by'\by)}{(p_\bgamma+3)\tFo (n/2,1;(
%p_\bgamma+3)/2;\by'\bP_\bgamma \by\big/\by'\by)}\,\bX_\text{new}\hat\bbeta^\bgamma\,,
%\end{eqnarray*}
If $\bX_\text{new}$ is a $q\times p$ matrix containing $q$ {\em new} values of
the explanatory variables for which we would like to predict the corresponding response
$\by_\text{new}$, the Bayesian predictor of $\by_\text{new}$ is given by
\begin{eqnarray*}
\hat\by_\text{new}^\bgamma
&=&\E\left[\by_\text{new}|\bX_\text{new},\bX,\bgamma,\by\right]\\
&=&2\,\frac{\tFo (n/2,2;(p_\bgamma+3)/2+1;\by'\bP_\bgamma \by\big/\by'\by)}{(p_\bgamma+3)\tFo (n/2,1;(
p_\bgamma+3)/2;\by'\bP_\bgamma \by\big/\by'\by)}\,\bX_\text{new}\hat\bbeta^\bgamma\,.
\end{eqnarray*}
Similarly, the Bayesian model averaging predictor of $\by_\text{new}$ is given by
\begin{eqnarray}
\hat\by_\text{new}&=&\E\left[\by_\text{new}|\bX_\text{new},\bX,\by\right]\label{eq:bambam}\\
&=&2\,\frac{\sum_{\bgamma\in\Gamma} \tFo (n/2,2;(p_\bgamma+3)/2+1;\by'\bP_\bgamma \by\big/\by'\by)/\left
[(p_\bgamma+1)(p_\bgamma+3)\right]}
{\sum_{\bgamma\in\Gamma} \tFo(n/2,1;(p_\bgamma+3)/2;\by'\bP_\bgamma \by\big/\by'\by)
/(p_\bgamma+1)}\,\bX_\text{new}\hat\bbeta^\bgamma\,.\nonumber
\end{eqnarray}
This numerical simplification in the derivation of Bayesian estimates and predictors is found in 
\cite{Liang:Molina:Clyde:Berger:2008} and exploited further in \cite{Bottolo:Richardson:2010}.
Note also that \cite{Guo:Speckman:2009} have furthermore established the consistency of the Bayes
factors based on such priors.

In contrast with this proposal, the prior of \cite{Liang:Molina:Clyde:Berger:2008} depends on a tuning parameter $a$.
% The choice of this hyperparameter $a$ is sensitive and its influence is unfortunately non-vanishing against an increase of
% the number of observations $n$, since $g$ has a significant influence on the Bayesian analysis of the linear model, as discussed earlier. 
Despite that, there also exist arguments to support this prior modelling, including the important issue of invariance under
location-scale transforms. As seen in the above formulae, the Jeffreys prior associated to model $\mathcal{M}_\bgamma$ ensure scale
invariance but not location invariance. In order to ensure location invariance for model $\mathcal{M}_\bgamma$,
it would be necessary to center the observation variable $y$ as well as the dependent variables $X$. Obviously, this centering of the data is 
completely unjustified from a Bayesian perspective and further it creates artificial correlations between observations. However it could be argued
that the lack of location invariance only pertains to quite specific and somehow artificial situations and that it is negligible in most situations. 
We will return to this point in the comparison section.

A location scale alternative consists in using the prior of \cite{Liang:Molina:Clyde:Berger:2008} with $a=2$
and excluding the null model from the competitors.
This prior leads  to the model posterior probability
\begin{equation}
\pi(\bgamma|\bX,\by)\propto \frac{\tFo((n-1)/2,1;(p_\bgamma+2)/2;(\by-\bar\by)'\bP_\bgamma
(\by-\bar\by) \big/(\by-\bar\by)'(\by-\bar\by))}{p_\bgamma}\,.
\label{eq:mpost2}
\end{equation}

Equations \eqref{eq:mpost1} and \eqref{eq:mpost2} are similar. However, in the last part of \eqref{eq:mpost2},
$\by$ is centered, ensuring the location invariance of the selection procedure.

%The possible advantages
%of this prior distribution are twofold: it is location invariant and not depending of any 
%tuning parameters.

\section{Numerical comparisons}\label{sec:simu}
We present here the results of numerical experiments aiming at comparing the behavior
of Bayesian variable selection and of some (non-Bayesian) popular regularization methods in regression, 
when considered from a variable selection point of view: The regularization methods that we consider are the Lasso, 
the Dantizg selector, and elastic net, described in Section \ref{lasso}. The Bayesian variable selection procedures
we consider oppose strategies for selecting the hyperparameter $g$ in Zellner's $g$-priors:
We include in this comparison the intrinsic prior \citep{Casella:Moreno:2006} which is 
another default objective prior for the non informative setting that does not require any tuning parameters and is
also invariant under location and scale changes. All procedure under comparison are described in Table \ref{tab1}.
We have also included in this comparison the highly standard AIC and BIC penalized likelihood criteria.
Moreover, we will refer to the performances of an ORACLE procedure that assumes the true model is known
and that estimate the regression coefficients with the least squares method.

\subsection{Regularization methods}
\label{lasso}

\begin{description}
  \item[\emph{1) The Lasso}:] Introduced by \cite{Tibshirani:1996}, the Lasso is a shrinkage method for linear regression. 
It is defined as the solution to the 
following 
$\ell_1$ penalized least squares optimization problem 
  $$
{\hat{\boldsymbol{\beta}}}_{\mbox{Lasso}}=\arg\min_{\boldsymbol{\beta}}||\mathbf{y}-
X\boldsymbol{\beta}||_2^2 +\lambda\sum_{j = 1}^p|\beta_j|,
$$
where $\lambda$ is a positive tuning parameter.
\item[\emph{2) The Dantzig Selector}:]
\cite{Candes:Tao:2007} introduced the Dantzig Selector as an alternative
to the Lasso. The Dantzig Selector is the solution to the optimization problem
$$
\min_{\beta\in\mathbb{R}^{p}} \|\beta\|_{1} \hspace*{3mm}\mbox{subject to}\hspace*{3mm}\
\|\mathbf{X}^{t}(\mathbf{y} - \mathbf{X}\beta)\|_{\infty}\leq \lambda,
$$
where $\lambda$ is a positive tuning parameter. The constraint
$\|\mathbf{X}^{t}(\mathbf{y} - \mathbf{X}\beta)\|_{\infty}\leq \lambda$ can be viewed as a relaxation of the normal equation in the classical linear regression.
  \item[\emph{3) The Elastic Net (Enet)}:]
The Lasso has at least two limitations: a) Lasso
does not encourage grouped selection in the presence of high
correlated covariates and b) for the $p>n$ case Lasso can select at most
$n$ covariates. To overcome these limitations, \cite{Zou:Hastie:2005} 
proposed an elastic net that combines both ridge $\ell_2$ and Lasso
$\ell_1$ penalties, i.e.
$$
{\hat{\boldsymbol{\beta}}}_{\mbox{Enet}}=\arg\min_{\boldsymbol{\beta}}||\mathbf{y}-
X\boldsymbol{\beta}||_2^2 + \lambda\sum_{j = 1}^p |\beta_{j}|+ \mu\sum_{j = 1}^p \beta_{j}^{2},
$$
where $\lambda$ and $\mu$ are two positive tuning parameters.
%  \item[\emph{4) The Elastic Corr-Net (CNET)}:] CNET \citep{Elanbari:Mkhadri:2008} 
% is a modification of Enet in which the ridge penalty term is replaced by
% the correlation based penalty term $P_c(\boldsymbol{\beta})$ defined by
% $$
% P_{c}(\boldsymbol{\beta)} = \sum_{j=1}^{p-1} \sum_{j > i} \left \{
% \frac{(\beta_{i} - \beta_{j})^{2}}{1 - \rho_{ij}} + \frac{(\beta_{i}
% + \beta_{j})^{2}}{1 + \rho_{ij}}\right\},
% $$
% where $\rho_{ij}=\mathbf{x}_i^t\mathbf{x}_j$ denotes the (empirical)
% correlation between the $i$th and the $j$th predictors. The
% correlation based penalty $P_c(\boldsymbol{\beta})$, introduced by Tutz and
% Ulbricht (2009), is expected to encourage a grouping effect for highly
% correlated variables.
\end{description}

\subsection{Numerical experiments on simulated datasets}
We have designed six different simulated datasets as benchmarks chosen as follows:

\begin{enumerate}
\item Example 1 (\textbf{sparse uncorrelated design}) corresponds to an uncorrelated covariate setting ($\rho=0$),
with $p=10$ predictors and where the components of $\mathbf{x}_{i}$ ($i=1,\ldots,10$) are iid
$\mathcal{N}_1(0,1)$ realizations. The response is simulated as
$$
\by\sim\mathcal{N}_n(2+\bx_2+2\bx_3-2\bx_6-1.5\bx_7,I_n)\,.
$$

\item Example 2 (\textbf{sparse correlated design}) corresponds to a correlated case ($\rho=0.9$), with $p=10$ predictors
and $\bx_{i}=(\mathbf{z}_{i}+3\bz_{11})/\sqrt{10}$, for $i=1,2$,
$\bx_{i}=(\bz_{i}+3\bz_{12})/\sqrt{10}$, for $i=3,4,5$,
and $\bx_{i}=(\bz_{i}+3\bz_{13})/\sqrt{10}$ for $i=6,\ldots,10$, the
components of $\mathbf{z}_{i}$ ($i=1,\ldots,13$) being iid
$\mathcal{N}_1(0,1)$ realizations. The use of
common terms in the $\bx_{i}$'s obviously induces a correlation among those $\bx_{i}$'s:
the correlation between variables $\bx_1$ and $\bx_2$ is 0.9, as for the variables ($\bx_3$, $\bx_4$ and $\bx_5$), and for the 
variables ($\bx_6$, $\bx_7$, $\bx_8$, $\bx_9$ and $\bx_{10}$). There is no correlation between those three groups of variables.
The response is simulated as
$$
\by\sim\mathcal{N}_n(2+\bx_2+2\bx_3-2\bx_6-1.5\bx_7,I_n)\,.
$$

\item Example 3 (\textbf{sparse noisy correlated design}) involves $p=8$ predictors. Those variables are generated using a
multivariate Gaussian distribution with correlations
$$
\rho(\mathbf{x}_{i},\mathbf{x}_{j}) = 0.5^{|i - j|}\,.
$$
The response is simulated as
$$
\mathbf{y}\sim\mathcal{N}_{n}(3\bx_1+1.5\bx_2+2\bx_5,9I_n)\,.
$$

%\item Example $4$ (\textbf{saturated noisy correlated design})
%is the same as Example $3$, except that the response is simulated as
%$$
%\mathbf{y}\sim\mathcal{N}_{n}\left(0.85\sum_{i=1}^8\bx_i,9I_n\right)\,.
%$$

\item Example $4$ (\textbf{saturated correlated design})
is the same as Example $4$, except that the response is simulated as
$$
\mathbf{y}\sim\mathcal{N}_{n}\left(0.85\sum_{i=1}^8\bx_i,I_n\right)\,.
$$

\item Example 5 involves $p=9$ predictors. Those variables are generated using a
multivariate Gaussian distribution with correlations
$$
\rho(\mathbf{x}_{i},\mathbf{x}_{j}) = 0.7^{|i - j|}\,.
$$
The response is simulated as
$$
\mathbf{y}\sim\mathcal{N}_{n}(2\bx_2-3\bx_4,I_n)\,.
$$

\item Example 6 (\textbf{null model}) involves $p=8$ predictors. Those variables are generated using a
multivariate Gaussian distribution with correlations
$$
\rho(\mathbf{x}_{i},\mathbf{x}_{j}) = 0.5^{|i - j|}\,.
$$
The response is simulated as
$$
\mathbf{y}\sim\mathcal{N}_{n}(2, 4I_n)\,.
$$

\end{enumerate}

Each dataset consists of a training set of size $n=15$, on which the regression model has been fitted and a test
set $T$ of size $n_T=200$ for assessing performances. Tuning parameters in the Lasso, the Dantzig
selector (DZ), and the elastic net (ENET) have been selected by
minimizing the cross-validation prediction error through leave-one-out. 
For each example, $100$ independent datasets have been simulated. We use three measures
of performances:

\begin{enumerate}
\item The root mean squared error (MSE)

\centerline{$
\mbox{MSE}_{y}=\sqrt{{\sum_{i=1}^{n_T} (y_i - \hat y_i)^2}\big/n_T}\,,
$}

$\hat y_i$ being the prediction of $y_i$ in the test set;

\item HITS: the number of correctly identified influential variables;

\item FP (False Positives): the number of non-influential variables declared as influential.
\end{enumerate}

Using those six different datasets as benchmarks, we compare the variable selection methods listed in Table \ref{tab1}.
The performances of the above selection methods are summarized in Tables \ref{example1.1}--\ref{example6.2}.
In the Bayesian approaches, the set of variables is naturally selected according to the maximum posterior probability
$\pi(\bgamma|\bX,\by)$ and the predictive is obtained via the Bayesian model averaging predictors.
%Note that we also ran a test about the modified behavior of our approach (NIMS, which stands for non-informative mixture selection) 
%when the response variable is drifted by $20$, i.e.~using $\by + 20$ instead of $\by$. As we expected, a change 
%in the location of the response does not affect at all the statistical performances of the analysis obtained with our approach. 

\begin{table}[htbp]
\label{methods}
\begin{center}
\begin{tabular}{|l|l|}
  \hline
  AIC     & Akaike Information Criterion \\
  BIC     & Bayesian Information Criterion \\
  \hline
  \hline
  BRIC    & g prior with $g=\max(n,p^2)$ \citep{Fernandez:Ley:Steel:2001} \\
  EB-L    & Local EB estimate of $g$ in $g$-prior \citep{Cui:George:2008}\\
  EB-G    & Global EB estimate of $g$ in $g$-prior \citep{Cui:George:2008} \\
  ZS-N    & Base model in Bayes factor taken as the null model \citep{Liang:Molina:Clyde:Berger:2008} \\
  ZS-F    & Base model in Bayes factor taken as the full model \citep{Liang:Molina:Clyde:Berger:2008} \\
  OVS     & Objective variable selection using the intrinsic prior \citep{Casella:Moreno:2006} \\
  HG-$3$  & Hyper-g prior with $a = 3$ \citep{Liang:Molina:Clyde:Berger:2008} \\
  HG-$4$  & Hyper-g prior with $a = 4$ \citep{Liang:Molina:Clyde:Berger:2008} \\
  \hline
  \hline
  HG-$2$  & Hyper-g prior with $a = 2$ \citep{Liang:Molina:Clyde:Berger:2008}, null model excluded \\
  NIMS    & Jeffreys prior on the non-invariant model \\
  \hline
  \hline
  LASSO   & Lasso \citep{Tibshirani:1996} \\
  DZ      & The Dantzig Selector \citep{Candes:Tao:2007} \\
  ENET    & The elastic-net \citep{Zou:Hastie:2005} \\
  \hline
 \end{tabular}
\caption{\label{tab1}Accronyms and description for the variable selection methods compared in the numerical experiment.
(The block separate the methods by their nature.}
\end{center}
\end{table}

\vs In this numerical experiment, the Bayesian procedures are clearly much more parsimonious than the regularization
procedures in that they almost always avoid overfitting. 
In all examples, the false positive rate FP is smaller for the Bayesian
solutions than  for the regularization methods. 
Except for the ZS-F and OVS scenarios 
which behave slightly worse than the others, all the Bayesian
procedures tested here produce the same selection of predictors. 
It seems that ZS-F has a slight tendency to select too many 
variables. The performances of OVS are somewhat disappointing and this
procedure seems to have a tendency to be too parsimonious. 
From a predictive viewpoint, computing the MSE by model averaging, Bayesian approaches also perform better than regularization
approaches except for the saturated correlated example (Example 4). We further note that
the classical selection procedures based on AIC and BIC do not easily reject variables and 
are thus slightly worse than Bayesian and regularization procedures (a fact not surprising
for AIC).  In all examples, the NIMS  and HG-2 approaches lead to optimal
performances in that they select the right covariates and only the
right covariates, while achieving close to the minimal root mean squared
error compared with all the other Bayesian solutions we considered.
They also do almost systematically better than BIC and
AIC. 
%The case of Example 4 is rather extreme
%due to the large noise factor, but the standard regularization procedures
%manage to reduce the MSE in this case by close to 10\% compared with the
%Bayesian procedures.

A global remark about this coparison is that all Bayesian procedures have a very similar MSE
and thus that they all correspond to the same regularization effect, except
for OVS which does systematically worse. However it is important to notice that the MSE for
OVS has not been computed by model averaging, but by using the best model. 
Otherwise, it would be hazardous to recommend one of the priors from those simulations
since there is no sensitive difference between them from both selection and prediction points of view. 
\begin{table}[htbp]
\begin{center}
\begin{tabular}{|l|ccc|}
  \hline
            & $MSE_{y}$ & HITS & FP \\
  \hline
  ORACLE    & $1.24 (0.02)$ & $4.00 (0.00)$ & $0.00 (0.00)$ \\
  \hline
  \hline
  AIC       & $1.75 (0.08)$ & $3.94 (0.02)$ & $2.78 (0.17)$ \\
  BIC       & $1.69 (0.08)$ & $3.90 (0.03)$ & $2.29 (0.17)$ \\
  \hline
  \hline
  BRIC      & $1.43 (0.04)$ & $3.75 (0.05)$ & $0.65 (0.09)$ \\
  EB-L      & $1.46 (0.04)$ & $3.80 (0.04)$ & $0.66 (0.09)$ \\
  EB-G      & $1.45 (0.04)$ & $3.78 (0.04)$ & $0.65 (0.09)$ \\
  ZS-N      & $1.44 (0.03)$ & $3.78 (0.04)$ & $0.65 (0.09)$ \\
  ZS-F      & $1.49 (0.03)$ & $3.90 (0.03)$ & $1.73 (0.14)$ \\
  OVS       & $1.52 (0.06)$ & $3.63 (0.06)$ & $0.54 (0.09)$ \\
  HG-$3$    & $1.49 (0.04)$ & $3.75 (0.05)$ & $0.55 (0.09)$ \\
  HG-$4$    & $1.57 (0.04)$ & $3.65 (0.05)$ & $0.54 (0.08)$ \\
  \hline
  \hline
  HG-$2$    & $1.50 (0.04)$ & $3.75 (0.05)$ & $0.59 (0.09)$ \\
  NIMS       & $1.45 (0.03)$ & $3.75 (0.05)$ & $0.57 (0.08)$\\
  \hline
  \hline
  LASSO     & $1.67 (0.05)$ & $3.89 (0.03)$ & $2.68 (0.20)$ \\
  DZ        & $1.66 (0.06)$ & $3.72 (0.07)$ & $2.41 (0.15)$ \\
  ENET      & $1.72 (0.05)$ & $3.89 (0.04)$ & $2.79 (0.29)$ \\
  \hline
\end{tabular}
\caption{\label{example1.1} Example 1: Mean of MSE, HITS and FP. The numbers between parentheses are the corresponding standard errors.}
\end{center}
\end{table}

\begin{table}[htbp]
\begin{center}
\begin{tabular}{|l|cccccccccc|}
  \hline
  Variables & $1$ & $2$     & $3$ & $4$    & $5$    & $6$ & $7$ & $8$    & $9$    & $10$ \\
  \hline
  AIC     & $0.47$ & $0.95$ & $1.00$ & $0.45$ & $0.44$ & $0.99$ & $1.00$ & $0.46$ & $0.52$ & $0.44$ \\
  BIC     & $0.41$ & $0.91$ & $1.00$ & $0.38$ & $0.40$ & $0.99$ & $1.00$ & $0.32$ & $0.44$ & $0.34$ \\
  \hline
  \hline
  BRIC    & $0.18$ & $0.77$ & $1.00$ & $0.10$ & $0.11$ & $0.99$ & $0.99$ & $0.07$ & $0.10$ & $0.09$ \\
  EB-L    & $0.17$ & $0.81$ & $1.00$ & $0.11$ & $0.11$ & $0.99$ & $1.00$ & $0.07$ & $0.11$ & $0.09$ \\
  EB-G    & $0.17$ & $0.79$ & $1.00$ & $0.11$ & $0.11$ & $0.99$ & $1.00$ & $0.07$ & $0.10$ & $0.09$ \\
  ZS-N    & $0.17$ & $0.79$ & $1.00$ & $0.11$ & $0.11$ & $0.99$ & $1.00$ & $0.07$ & $0.10$ & $0.09$ \\
  ZS-F    & $0.34$ & $0.90$ & $1.00$ & $0.29$ & $0.33$ & $1.00$ & $1.00$ & $0.20$ & $0.33$ & $0.24$ \\
  OVS     & $0.14$ & $0.72$ & $0.98$ & $0.07$ & $0.08$ & $0.97$ & $0.96$ & $0.08$ & $0.10$ & $0.07$ \\
  HG-$3$  & $0.17$ & $0.77$ & $1.00$ & $0.11$ & $0.10$ & $0.99$ & $0.99$ & $0.07$ & $0.09$ & $0.08$ \\
  HG-$4$  & $0.15$ & $0.77$ & $1.00$ & $0.10$ & $0.08$ & $0.99$ & $0.99$ & $0.07$ & $0.08$ & $0.07$ \\
  \hline
  \hline
  HG-$2$  & $0.10$ & $0.83$ & $0.99$ & $0.07$ & $0.16$ & $0.98$ & $0.95$ & $0.13$ & $0.06$ & $0.07$ \\
  NIMS     & $0.15$ & $0.77$ & $1.00$ & $0.09$ & $0.09$ & $0.99$ & $0.99$ & $0.06$ & $0.10$ & $0.08$ \\
  \hline
  \hline
  LASSO   & $0.49$ & $0.91$ & $1.00$ & $0.41$ & $0.45$ & $0.98$ & $1.00$ & $0.49$ & $0.47$ & $0.37$ \\
  DZ      & $0.42$ & $0.84$ & $0.96$ & $0.41$ & $0.47$ & $0.97$ & $0.95$ & $0.38$ & $0.37$ & $0.36$ \\
  ENET    & $0.45$ & $0.93$ & $1.00$ & $0.45$ & $0.43$ & $0.99$ & $0.97$ & $0.52$ & $0.44$ & $0.50$ \\
  \hline
\end{tabular}
\caption{\label{example1.2} Example 1: Relative frequencies of the selected variables for methods under comparison.}
\end{center}
\end{table}

\begin{table}[htbp]
\begin{center}
\begin{tabular}{|l|ccc|}
  \hline
            & $MSE_{y}$ & HITS & FP \\
  \hline
  ORACLE    & $1.19 (0.01)$ & $4.00 (0.00)$ & $0.00 (0.00)$ \\
  \hline
  \hline
  AIC       & $1.81 (0.06)$ & $3.12 (0.08)$ & $2.75 (0.16)$ \\
  BIC       & $1.76 (0.05)$ & $2.97 (0.09)$ & $2.39 (0.16)$ \\
  \hline
  \hline
  BRIC      & $1.46 (0.02)$ & $2.44 (0.10)$ & $0.99 (0.10)$ \\
  EB-L      & $1.45 (0.02)$ & $2.43 (0.10)$ & $1.03 (0.10)$ \\
  EB-G      & $1.45 (0.02)$ & $2.42 (0.10)$ & $0.95 (0.10)$ \\
  ZS-N      & $1.45 (0.02)$ & $2.43 (0.10)$ & $1.03 (0.10)$ \\
  ZS-F      & $1.42 (0.02)$ & $2.97 (0.08)$ & $2.18 (0.10)$ \\
  OVS       & $1.71 (0.04)$ & $2.16 (0.11)$ & $1.09 (0.09)$ \\
  HG-$3$    & $1.45 (0.02)$ & $2.32 (0.11)$ & $0.96 (0.10)$ \\
  HG-$4$    & $1.45 (0.02)$ & $2.35 (0.10)$ & $0.86 (0.09)$ \\
  \hline
  \hline
  HG-$2$    & $1.52 (0.04)$ & $2.35 (0.10)$ & $0.81 (0.09)$ \\
  NIMS       & $1.45 (0.02)$ & $2.42 (0.10)$ & $0.96 (0.09)$\\
  \hline
  \hline
  LASSO     & $1.66 (0.05)$ & $3.35 (0.09)$ & $2.95 (0.15)$ \\
  DZ        & $1.59 (0.03)$ & $2.83 (0.09)$ & $2.23 (0.10)$ \\
  ENET      & $1.50 (0.03)$ & $3.70 (0.07)$ & $4.36 (0.17)$ \\
  \hline
\end{tabular}
\caption{\label{example2.1} Example 2: Mean of MSE, HITS and FP. The numbers between parentheses are the corresponding standard errors.}
\end{center}
\end{table}

\begin{table}[htbp]
\begin{center}
\begin{tabular}{|l|cccccccccc|}
  \hline
  Variables & $1$ & $2$     & $3$ & $4$    & $5$    & $6$ & $7$ & $8$    & $9$    & $10$ \\
  \hline
  \hline
  AIC     & $0.46$ & $0.79$ & $0.88$ & $0.44$ & $0.46$ & $0.78$ & $0.67$ & $0.52$ & $0.48$ & $0.39$ \\
  BIC     & $0.41$ & $0.71$ & $0.86$ & $0.43$ & $0.33$ & $0.77$ & $0.63$ & $0.42$ & $0.45$ & $0.35$ \\
  \hline
  \hline
  BRIC    & $0.21$ & $0.60$ & $0.80$ & $0.17$ & $0.13$ & $0.65$ & $0.39$ & $0.18$ & $0.18$ & $0.12$ \\
  EB-L    & $0.22$ & $0.59$ & $0.80$ & $0.17$ & $0.14$ & $0.66$ & $0.38$ & $0.19$ & $0.19$ & $0.12$ \\
  EB-G    & $0.21$ & $0.59$ & $0.81$ & $0.16$ & $0.13$ & $0.65$ & $0.37$ & $0.19$ & $0.16$ & $0.10$ \\
  ZS-N    & $0.22$ & $0.59$ & $0.80$ & $0.17$ & $0.14$ & $0.66$ & $0.38$ & $0.19$ & $0.19$ & $0.12$ \\
  ZS-F    & $0.40$ & $0.72$ & $0.84$ & $0.37$ & $0.31$ & $0.79$ & $0.62$ & $0.38$ & $0.41$ & $0.31$ \\
  OVS     & $0.23$ & $0.44$ & $0.74$ & $0.17$ & $0.23$ & $0.62$ & $0.36$ & $0.19$ & $0.18$ & $0.09$ \\
  HG-$3$  & $0.21$ & $0.54$ & $0.80$ & $0.16$ & $0.13$ & $0.63$ & $0.35$ & $0.18$ & $0.18$ & $0.10$ \\
  HG-$4$  & $0.18$ & $0.56$ & $0.81$ & $0.15$ & $0.11$ & $0.63$ & $0.35$ & $0.17$ & $0.17$ & $0.08$ \\
  \hline
  \hline
  HG-$2$  & $0.22$ & $0.60$ & $0.78$ & $0.16$ & $0.13$ & $0.59$ & $0.42$ & $0.10$ & $0.15$ & $0.11$ \\
  NIMS     & $0.19$ & $0.59$ & $0.80$ & $0.16$ & $0.14$ & $0.66$ & $0.37$ & $0.19$ & $0.18$ & $0.10$ \\
  \hline
  \hline
  LASSO   & $0.47$ & $0.77$ & $0.90$ & $0.53$ & $0.40$ & $0.89$ & $0.79$ & $0.57$ & $0.55$ & $0.43$ \\
  DZ      & $0.40$ & $0.65$ & $0.79$ & $0.46$ & $0.37$ & $0.76$ & $0.63$ & $0.32$ & $0.36$ & $0.32$ \\
  ENET    & $0.68$ & $0.85$ & $0.97$ & $0.74$ & $0.74$ & $0.96$ & $0.92$ & $0.76$ & $0.75$ & $0.69$ \\
  \hline
\end{tabular}
\caption{\label{example2.2} Example 2: Relative frequencies of the selected variables for methods under comparison.}
\end{center}
\end{table}

\begin{table}[htbp]
\begin{center}
\begin{tabular}{|l|ccc|}
  \hline
            & $MSE_{y}$ & HITS & FP \\
 \hline
  ORACLE    & $3.31 (0.03)$ & $3.00 (0.00)$ & $0.00 (0.00)$ \\
  \hline
  \hline
  AIC       & $4.32 (0.09)$ & $2.11 (0.07)$ & $2.06 (0.14)$ \\
  BIC       & $4.24 (0.08)$ & $1.97 (0.07)$ & $1.68 (0.14)$ \\
  \hline
  \hline
  BRIC      & $4.07 (0.07)$ & $1.66 (0.07)$ & $0.53 (0.08)$ \\
  EB-L      & $4.06 (0.06)$ & $1.84 (0.07)$ & $0.79 (0.09)$ \\
  EB-G      & $4.07 (0.07)$ & $1.88 (0.07)$ & $0.83 (0.09)$ \\
  ZS-N      & $4.01 (0.06)$ & $1.81 (0.07)$ & $0.76 (0.09)$ \\
  ZS-F      & $4.04 (0.07)$ & $2.10 (0.07)$ & $1.26 (0.11)$ \\
  OVS       & $4.27 (0.09)$ & $1.78 (0.07)$ & $0.64 (0.09)$ \\
  HG-$3$    & $4.05 (0.06)$ & $1.81 (0.07)$ & $0.77 (0.09)$ \\
  HG-$4$    & $4.08 (0.06)$ & $1.84 (0.07)$ & $0.78 (0.09)$ \\
  \hline
  \hline
  HG-$2$    & $3.98 (0.05)$ & $1.80 (0.08)$ & $0.73 (0.10)$ \\
  NIMS       & $3.99 (0.06)$ & $1.83 (0.07)$ & $0.77 (0.09)$\\
  \hline
  \hline
  LASSO     & $4.03 (0.06)$ & $2.33 (0.07)$ & $1.61 (0.16)$ \\
  DZ        & $4.32 (0.10)$ & $2.20 (0.11)$ & $2.06 (0.16)$ \\
  ENET      & $4.13 (0.06)$ & $2.38 (0.06)$ & $2.04 (0.16)$ \\
  \hline
\end{tabular}
\caption{\label{example3.1} Example 3: Mean of MSE, HITS and FP. The numbers between parentheses are the corresponding standard errors.}
\end{center}
\end{table}

\begin{table}[htbp]
\begin{center}
\begin{tabular}{|l|cccccccc|}
  \hline
  Variables & $1$ & $2$ & $3$ & $4$ & $5$ & $6$ & $7$ & $8$ \\
  \hline
  AIC     & $0.89$ & $0.52$ & $0.45$ & $0.43$ & $0.70$ & $0.36$ & $0.42$ & $0.40$ \\
  BIC     & $0.89$ & $0.44$ & $0.39$ & $0.36$ & $0.64$ & $0.30$ & $0.33$ & $0.30$ \\
  \hline
  \hline
  BRIC    & $0.82$ & $0.35$ & $0.09$ & $0.13$ & $0.49$ & $0.12$ & $0.08$ & $0.11$ \\
  EB-L    & $0.87$ & $0.38$ & $0.13$ & $0.19$ & $0.59$ & $0.18$ & $0.14$ & $0.15$ \\
  EB-G    & $0.89$ & $0.39$ & $0.15$ & $0.20$ & $0.60$ & $0.18$ & $0.14$ & $0.16$ \\
  ZS-N    & $0.87$ & $0.37$ & $0.13$ & $0.19$ & $0.57$ & $0.16$ & $0.13$ & $0.15$ \\
  ZS-F    & $0.92$ & $0.51$ & $0.23$ & $0.34$ & $0.67$ & $0.22$ & $0.24$ & $0.23$ \\
  OVS     & $0.86$ & $0.37$ & $0.12$ & $0.14$ & $0.55$ & $0.16$ & $0.08$ & $0.14$ \\
  HG-$3$  & $0.87$ & $0.38$ & $0.13$ & $0.19$ & $0.56$ & $0.16$ & $0.14$ & $0.15$ \\
  HG-$4$  & $0.88$ & $0.38$ & $0.13$ & $0.19$ & $0.58$ & $0.17$ & $0.14$ & $0.15$ \\
  \hline
  \hline
  HG-$2$  & $0.80$ & $0.46$ & $0.19$ & $0.17$ & $0.60$ & $0.21$ & $0.12$ & $0.15$ \\
  NIMS    & $0.87$ & $0.38$ & $0.12$ & $0.19$ & $0.58$ & $0.17$ & $0.14$ & $0.15$ \\
  \hline
  \hline
  LASSO   & $0.96$ & $0.70$ & $0.32$ & $0.40$ & $0.67$ & $0.29$ & $0.23$ & $0.37$ \\
  DZ      & $0.82$ & $0.71$ & $0.42$ & $0.47$ & $0.67$ & $0.47$ & $0.31$ & $0.39$ \\
  ENET    & $0.97$ & $0.71$ & $0.49$ & $0.50$ & $0.70$ & $0.40$ & $0.30$ & $0.35$ \\
  \hline
\end{tabular}
\caption{\label{example3.2} Example 3: Relative frequencies of the selected variables for methods under comparison.}
\end{center}
\end{table}

\begin{table}[htbp]
\begin{center}
\begin{tabular}{|l|ccc|}
  \hline
            & $MSE_{y}$ & HITS & FP \\
  \hline
  ORACLE    & $1.43 (0.03)$ & $8.00 (0.00)$ & $0.00 (0.00)$ \\
  \hline
  \hline
  AIC       & $1.60 (0.03)$ & $6.32 (0.11)$ & $0.00 (0.00) $ \\
  BIC       & $1.64 (0.03)$ & $5.99 (0.12)$ & $0.00 (0.00) $ \\
  \hline
  \hline
  BRIC      & $1.79 (0.04)$ & $4.35 (0.11)$ & $0.00 (0.00) $ \\
  EB-L      & $1.75 (0.04)$ & $4.39 (0.10)$ & $0.00 (0.00) $ \\
  EB-G      & $1.76 (0.04)$ & $4.34 (0.10)$ & $0.00 (0.00) $ \\
  ZS-N      & $1.74 (0.04)$ & $4.38 (0.10)$ & $0.00 (0.00) $ \\
  ZS-F      & $1.62 (0.04)$ & $5.37 (0.10)$ & $0.00 (0.00) $ \\
  OVS       & $2.22 (0.04)$ & $3.82 (0.10)$ & $0.00 (0.00) $ \\
  HG-$3$    & $1.76 (0.04)$ & $4.32 (0.10)$ & $0.00 (0.00) $ \\
  HG-$4$    & $1.78 (0.03)$ & $4.19 (0.09)$ & $0.00 (0.00) $ \\
  \hline
  \hline
  HG-$2$    & $1.77 (0.04)$ & $4.18 (0.11)$ & $0.00 (0.00) $ \\
  NIMS       & $1.75 (0.04)$ & $4.39 (0.10)$ & $0.00 (0.00) $\\
  \hline
  \hline
  LASSO     & $1.59 (0.04)$ & $7.13 (0.12)$ & $0.00 (0.00) $ \\
  DZ        & $1.56 (0.03)$ & $6.82 (0.11)$ & $0.00 (0.00) $ \\
  ENET      & $1.54 (0.03)$ & $7.53 (0.08)$ & $0.00 (0.00) $ \\
  \hline
\end{tabular}
\caption{\label{example4.1} Example 4: Mean of MSE, HITS and FP. The numbers between parentheses are the corresponding standard errors.}
\end{center}
\end{table}

\begin{table}[htbp]
\begin{center}
\begin{tabular}{|l|cccccccc|}
  \hline
  Variables & $1$ & $2$ & $3$ & $4$ & $5$ & $6$ & $7$ & $8$ \\
  \hline
  AIC     & $0.80$ & $0.81$ & $0.78$ & $0.75$ & $0.76$ & $0.86$ & $0.77$ & $0.79$ \\
  BIC     & $0.76$ & $0.76$ & $0.75$ & $0.72$ & $0.68$ & $0.83$ & $0.71$ & $0.78$ \\
  \hline
  \hline
  BRIC    & $0.45$ & $0.58$ & $0.50$ & $0.65$ & $0.54$ & $0.55$ & $0.48$ & $0.60$ \\
  EB-L    & $0.46$ & $0.57$ & $0.52$ & $0.67$ & $0.54$ & $0.54$ & $0.50$ & $0.59$ \\
  EB-G    & $0.45$ & $0.57$ & $0.52$ & $0.66$ & $0.54$ & $0.53$ & $0.48$ & $0.59$ \\
  ZS-N    & $0.46$ & $0.57$ & $0.52$ & $0.67$ & $0.54$ & $0.54$ & $0.49$ & $0.59$ \\
  ZS-F    & $0.62$ & $0.69$ & $0.60$ & $0.78$ & $0.65$ & $0.67$ & $0.62$ & $0.74$ \\
  OVS     & $0.38$ & $0.57$ & $0.45$ & $0.64$ & $0.40$ & $0.49$ & $0.44$ & $0.45$ \\
  HG-$3$  & $0.45$ & $0.57$ & $0.51$ & $0.67$ & $0.54$ & $0.53$ & $0.48$ & $0.57$ \\
  HG-$4$  & $0.44$ & $0.57$ & $0.48$ & $0.66$ & $0.51$ & $0.52$ & $0.45$ & $0.56$ \\
  \hline
  \hline
  HG-$2$  & $0.53$ & $0.56$ & $0.50$ & $0.50$ & $0.54$ & $0.55$ & $0.53$ & $0.47$ \\
  NIMS    & $0.46$ & $0.58$ & $0.51$ & $0.67$ & $0.54$ & $0.54$ & $0.50$ & $0.59$ \\
  \hline
  \hline
  LASSO   & $0.82$ & $0.90$ & $0.96$ & $0.92$ & $0.85$ & $0.91$ & $0.87$ & $0.90$ \\
  DZ      & $0.84$ & $0.85$ & $0.84$ & $0.82$ & $0.83$ & $0.91$ & $0.89$ & $0.84$ \\
  ENET    & $0.89$ & $0.93$ & $0.96$ & $0.97$ & $0.96$ & $0.93$ & $0.96$ & $0.93$ \\
  \hline
\end{tabular}
\caption{\label{example4.2} Example 4: Relative frequencies of the selected variables for methods under comparison.}
\end{center}
\end{table}

\begin{table}[htbp]
\begin{center}
\begin{tabular}{|l|ccc|}
  \hline
            & $MSE_{y}$ & HITS & FP \\
  \hline
  ORACLE    & $1.07 (0.09)$ & $2.00 (0.00)$ & $0.00 (0.00)$ \\
  \hline
  \hline
  AIC       & $1.48 (0.05)$ & $1.93 (0.02)$ & $2.88 (0.19)$ \\
  BIC       & $1.39 (0.04)$ & $1.94 (0.02)$ & $2.04 (0.18)$ \\
  \hline
  \hline
  BRIC      & $1.24 (0.02)$ & $1.93 (0.02)$ & $0.50 (0.09)$ \\
  EB-L      & $1.27 (0.02)$ & $1.93 (0.02)$ & $0.58 (0.10)$ \\
  EB-G      & $1.27 (0.02)$ & $1.93 (0.02)$ & $0.60 (0.10)$ \\
  ZS-N      & $1.26 (0.02)$ & $1.93 (0.02)$ & $0.57 (0.10)$ \\
  ZS-F      & $1.33 (0.03)$ & $1.94 (0.02)$ & $1.84 (0.14)$ \\
  OVS       & $1.32 (0.04)$ & $1.89 (0.03)$ & $0.76 (0.08)$ \\
  HG-$3$    & $1.28 (0.02)$ & $1.93 (0.02)$ & $0.53 (0.09)$ \\
  HG-$4$    & $1.30 (0.02)$ & $1.93 (0.02)$ & $0.54 (0.09)$ \\
  \hline
  \hline
  HG-$2$    & $1.25 (0.02)$ & $1.93 (0.02)$ & $0.36 (0.09)$ \\
  NIMS       & $1.22 (0.02)$ & $1.93 (0.02)$ & $0.57 (0.10)$\\
  \hline
  \hline
  LASSO     & $1.39 (0.03)$ & $1.99 (0.01)$ & $2.93 (0.21)$ \\
  DZ        & $1.36 (0.04)$ & $1.91 (0.03)$ & $2.70 (0.18)$ \\
  ENET      & $1.43 (0.03)$ & $1.96 (0.02)$ & $3.25 (0.20)$ \\
  \hline
\end{tabular}
\caption{\label{example5.1} Example 5: Mean of MSE, HITS and FP. The numbers between parentheses are the corresponding standard errors.}
\end{center}
\end{table}

\begin{table}[htbp]
\begin{center}
\begin{tabular}{|l|ccccccccc|}
  \hline
  Variables & $1$ & $2$     & $3$ & $4$    & $5$    & $6$ & $7$ & $8$    & $9$ \\
  \hline
  AIC     & $0.36$ & $0.94$ & $0.47$ & $0.99$ & $0.35$ & $0.36$ & $0.34$ & $0.53$ & $0.47$  \\
  BIC     & $0.30$ & $0.94$ & $0.38$ & $1.00$ & $0.26$ & $0.24$ & $0.22$ & $0.35$ & $0.29$  \\
  \hline
  \hline
  BRIC    & $0.10$ & $0.94$ & $0.09$ & $1.00$ & $0.10$ & $0.03$ & $0.05$ & $0.08$ & $0.05$  \\
  EB-L    & $0.10$ & $0.93$ & $0.14$ & $1.00$ & $0.11$ & $0.04$ & $0.05$ & $0.08$ & $0.06$  \\
  EB-G    & $0.11$ & $0.93$ & $0.14$ & $1.00$ & $0.11$ & $0.04$ & $0.05$ & $0.08$ & $0.07$  \\
  ZS-N    & $0.10$ & $0.93$ & $0.13$ & $1.00$ & $0.11$ & $0.04$ & $0.05$ & $0.08$ & $0.06$  \\
  ZS-F    & $0.29$ & $0.94$ & $0.32$ & $1.00$ & $0.23$ & $0.22$ & $0.19$ & $0.31$ & $0.28$  \\
  OVS     & $0.16$ & $0.92$ & $0.10$ & $0.97$ & $0.15$ & $0.07$ & $0.09$ & $0.11$ & $0.08$  \\
  HG-$3$  & $0.10$ & $0.93$ & $0.11$ & $1.00$ & $0.11$ & $0.03$ & $0.04$ & $0.08$ & $0.06$  \\
  HG-$4$  & $0.10$ & $0.93$ & $0.12$ & $1.00$ & $0.11$ & $0.03$ & $0.04$ & $0.08$ & $0.06$  \\
  \hline
  \hline
  HG-$2$  & $0.08$ & $0.95$ & $0.07$ & $1.00$ & $0.04$ & $0.03$ & $0.02$ & $0.06$ & $0.06$ \\
  NIMS     & $0.06$ & $0.97$ & $0.10$ & $1.00$ & $0.11$ & $0.08$ & $0.05$ & $0.08$ & $0.08$  \\
  \hline
  \hline
  LASSO   & $0.51$ & $0.99$ & $0.35$ & $1.00$ & $0.47$ & $0.38$ & $0.37$ & $0.41$ & $0.44$  \\
  DZ      & $0.50$ & $0.93$ & $0.32$ & $0.98$ & $0.42$ & $0.45$ & $0.26$ & $0.32$ & $0.43$  \\
  ENET    & $0.52$ & $0.96$ & $0.37$ & $1.00$ & $0.55$ & $0.44$ & $0.43$ & $0.50$ & $0.44$  \\
  \hline
\end{tabular}
\caption{\label{example5.2} Example 5: Relative frequencies of the selected variables for methods under comparison.}
\end{center}
\end{table}

\begin{table}[htbp]
\begin{center}
\begin{tabular}{|l|cc|}
  \hline
            & $MSE_{y}$  & FP \\
 \hline
  ORACLE    & $1.99 (0.01)$ & $0.00 (0.00)$ \\
  \hline
  \hline
  AIC       & $2.80 (0.07)$ & $3.16 (0.21)$ \\
  BIC       & $2.62 (0.06)$ & $2.24 (0.19)$ \\
  \hline
  \hline
  BRIC      & $2.19 (0.02)$ & $0.59 (0.11)$ \\
  EB-L      & $2.12 (0.02)$ & $2.87 (0.15)$ \\
  EB-G      & $2.11 (0.02)$ & $1.54 (0.19)$ \\
  ZS-N      & $2.26 (0.02)$ & $1.02 (0.17)$ \\
  ZS-F      & $2.31 (0.03)$ & $2.51 (0.17)$ \\
  OVS       & $2.57 (0.06)$ & $2.10 (0.17)$ \\
  HG-$3$    & $2.13 (0.02)$ & $2.18 (0.18)$ \\
  HG-$4$    & $2.10 (0.01)$ & $2.54 (0.17)$ \\
  \hline
  \hline
  HG-$2$    & $2.16 (0.02)$ & $2.17 (0.15)$ \\
  NIMS      & $2.24 (0.02)$ & $0.99 (0.13)$\\
  \hline
  \hline
  LASSO     & $2.19 (0.04)$ & $1.79 (0.22)$ \\
  DZ        & $2.57 (0.05)$ & $2.49 (0.20)$ \\
  ENET      & $2.20 (0.04)$ & $2.23 (0.23)$ \\
  \hline
\end{tabular}
\caption{\label{example6.1} Example 6: Mean of MSE and FP. The numbers between parentheses are the corresponding standard errors.}
\end{center}
\end{table}

\begin{table}[htbp]
\begin{center}
\begin{tabular}{|l|cccccccc|}
  \hline
  Variables & $1$ & $2$ & $3$ & $4$ & $5$ & $6$ & $7$ & $8$ \\
  \hline
  AIC     & $0.38$ & $0.36$ & $0.31$ & $0.37$ & $0.49$ & $0.42$ & $0.41$ & $0.42$ \\
  BIC     & $0.26$ & $0.22$ & $0.23$ & $0.26$ & $0.31$ & $0.36$ & $0.33$ & $0.27$ \\
  \hline
  \hline
  BRIC    & $0.09$ & $0.04$ & $0.07$ & $0.08$ & $0.08$ & $0.09$ & $0.09$ & $0.05$ \\
  EB-L    & $0.37$ & $0.27$ & $0.28$ & $0.30$ & $0.43$ & $0.43$ & $0.38$ & $0.41$ \\
  EB-G    & $0.19$ & $0.12$ & $0.16$ & $0.16$ & $0.21$ & $0.27$ & $0.25$ & $0.18$ \\
  ZS-N    & $0.14$ & $0.07$ & $0.11$ & $0.10$ & $0.16$ & $0.16$ & $0.18$ & $0.10$ \\
  ZS-F    & $0.29$ & $0.27$ & $0.23$ & $0.28$ & $0.41$ & $0.38$ & $0.34$ & $0.31$ \\
  OVS     & $0.26$ & $0.26$ & $0.36$ & $0.23$ & $0.28$ & $0.26$ & $0.28$ & $0.17$ \\
  HG-$3$  & $0.27$ & $0.21$ & $0.20$ & $0.26$ & $0.32$ & $0.35$ & $0.30$ & $0.27$ \\
  HG-$4$  & $0.32$ & $0.25$ & $0.23$ & $0.29$ & $0.40$ & $0.38$ & $0.35$ & $0.32$ \\
  \hline
  \hline
  HG-$2$  & $0.25$ & $0.19$ & $0.23$ & $0.25$ & $0.31$ & $0.35$ & $0.32$ & $0.27$ \\
  NIMS    & $0.12$ & $0.06$ & $0.10$ & $0.11$ & $0.14$ & $0.17$ & $0.18$ & $0.11$ \\
  \hline
  \hline
  LASSO   & $0.22$ & $0.17$ & $0.23$ & $0.22$ & $0.24$ & $0.25$ & $0.29$ & $0.17$ \\
  DZ      & $0.23$ & $0.30$ & $0.17$ & $0.20$ & $0.30$ & $0.27$ & $0.25$ & $0.23$ \\
  ENET    & $0.30$ & $0.26$ & $0.27$ & $0.25$ & $0.28$ & $0.28$ & $0.33$ & $0.26$ \\
  \hline
\end{tabular}
\caption{\label{example6.2} Example 6: Relative frequencies of the selected variables for methods under comparison.}
\end{center}
\end{table}

\paragraph{Translating the data}
Since NIMS is not location invariant, it is important to measure the impact of adding 
a constant to all observations. As stressed by a reviewer, when this constant goes to infinity,
keeping $n$ fixed, the last argument of $\tFo$ in (\ref{eq:mpost1}) goes to one for all models.
Thus if the empirical mean is large relative to the regression sum of squares, the
data end up having little input in distinguishing between models.
In order to measure this possible negative impact of adding a large constant, 
we replace in Example $1$ $\mathbf{y}$ by $\mathbf{y} = \mathbf{y} + 10^{k}$ 
RSS (Regression Sum of Squares) for $k\in \{1, 2, 3\}$. The results derived from NIMS criterion are 
summarized in Tables \ref{example_n_15.1.1_scaled} and \ref{example_n_15.1.2_scaled}: as predicted,
the NIMS criterion tends to choose the null model as $k$ increases and the null model with no variable
is always selected when $k=3$. Therefore some prior assumption must be made about the magnitude of the intercept
when using NIMS. Otherwise, the criterion is over-parsimonious. If this is a possible case, we suggest using
instead the HG-2 approach.

\begin{table}[htbp]
\begin{center}
\begin{tabular}{|l|ccc|}
  \hline
            & $MSE_{y}$ & HITS & FP \\

  \hline
  $\mathbf{y} = \mathbf{y} + 10\times RSS$          & $3.41 (0.03)$ & $0.15 (0.04)$ & $0.00 (0.00)$ \\
  $\mathbf{y} = \mathbf{y} + 10^{2}\times RSS$      & $3.59 (0.03)$ & $0.01 (0.01)$ & $0.00 (0.00)$ \\
  $\mathbf{y} = \mathbf{y} + 10^{3}\times RSS$      & $3.59 (0.02)$ & $0.00 (0.00)$ & $0.00 (0.00)$ \\
  \hline
\end{tabular}
\caption{\label{example_n_15.1.1_scaled} Example 1: Mean of MSE, HITS and FP after replacing $\mathbf{y}$ by $\mathbf{y} = \mathbf{y} + 10^{k} RSS$ for $k \in \{1, 2, 3\}$. The numbers
between parentheses are the corresponding standard errors for the NIMS selection procedure.}
\end{center}
\end{table}

\begin{table}[htbp]
\begin{center}
\begin{tabular}{|l|cccccccccc|}
  \hline
  Variables & $1$ & $2$ & $3$ & $4$ & $5$ & $6$ & $7$ & $8$ & $9$ & $10$\\
  \hline
  $\mathbf{y} = \mathbf{y} + 10\times RSS$        & $0.00$ & $0.00$ & $0.09$ & $0.00$ & $0.00$ & $0.05$ & $0.01$ & $0.00$ & $0.00$ & $0.00$ \\
  $\mathbf{y} = \mathbf{y} + 10^{2}\times RSS$    & $0.00$ & $0.00$ & $0.01$ & $0.00$ & $0.00$ & $0.01$ & $0.00$ & $0.00$ & $0.00$ & $0.00$ \\
  $\mathbf{y} = \mathbf{y} + 10^{3}\times RSS$    & $0.00$ & $0.00$ & $0.00$ & $0.00$ & $0.00$ & $0.00$ & $0.00$ & $0.00$ & $0.00$ & $0.00$ \\
  \hline
\end{tabular}
\caption{\label{example_n_15.1.2_scaled} Example 1: Relative frequencies of the selected variables after replacing $\mathbf{y}$ by $\mathbf{y} = \mathbf{y} + 10^{k} RSS$ for $k \in \{1, 2, 3\}$.}
\end{center}
\end{table}

\subsection{Real datasets}
Two datasets considered in this section are associated with a moderate number of variables
against the number of observations.  

\paragraph{Body fat dataset} The body fat dataset has been first used by \cite{penrose:nelson:fisher:1985}.
The corresponding study aims at estimating the percentage of body fat from various body
circumference measurements observed on 252 men. The thirteen regressor variables are:

{ \small \begin{enumerate}
\item
age,
\item
weight (lbs),
\item
height (inches),
\item
neck circumference,
\item
chest circumference,
\item
abdomen $2$ circumference,
\item
hip circumference,
\item
thigh circumference,
\item
knee circumference,
\item
ankle circumference,
\item
biceps (extended) circumference,
\item
forearm circumference,
\item
%and 
wrist circumference.
\end{enumerate}}

In order to investigate the performances of the different methods, a dataset from \cite{penrose:nelson:fisher:1985}
has been split $25$ times into a training set of
$151$ observations and a test set of $101$ observations. Tuning parameters for the frequentist regularization
methods have been chosen by minimizing the (ten fold) cross-validated prediction error.

For this dataset, the Bayesian procedures we investigated are much
more parsimonious than the standard regularization procedures, as shown in Table
\ref{body1}. There is no variability in the prediction MSE.
(We stress that MSEs are computed by model averaging for the Bayesian procedures.) 
As in the simulation experiment, all Bayesian approaches are highly similar, except for
ZS-F which remains more open to incorporating the last two covariates.

\paragraph{Ozone data}
This second benchmark dataset is taken from \cite{Breiman:Friedman:1985}
and consists in daily measurements of the maximum ozone concentration and of eight meteorological variables near Los Angeles.
Those variables are:
{\small \begin{enumerate}
\item 
the daily ozone concentration (maximum one hour average, parts per million) at Upland, CA which is the response variable;
\item 
the Vandenburg 500 millibar pressure height (m);
\item 
the wind speed (mph) at Los Angeles International Airport (LAX);
\item 
the humidity (percent) at LAX;
\item 
the Sandburg Air Force Base temperature ($F^{o}$);
\item 
the inversion base height at LAX;
\item 
the inversion base temperature at LAX;
\item 
the Daggett Pressure gradient (mm Hg) from LAX to Daggett, CA;
\item 
the visibility (miles) at LAX.
\end{enumerate}}

The original Ozone database contains $366$ observations, of which $203$ are complete. Our study is made just on the complete observations. We split this dataset $25$ times into a training set of $101$ observations and a test set of $102$ observations.

\begin{figure}[h]
\begin{center}
 \includegraphics[width=16cm, height=10cm]{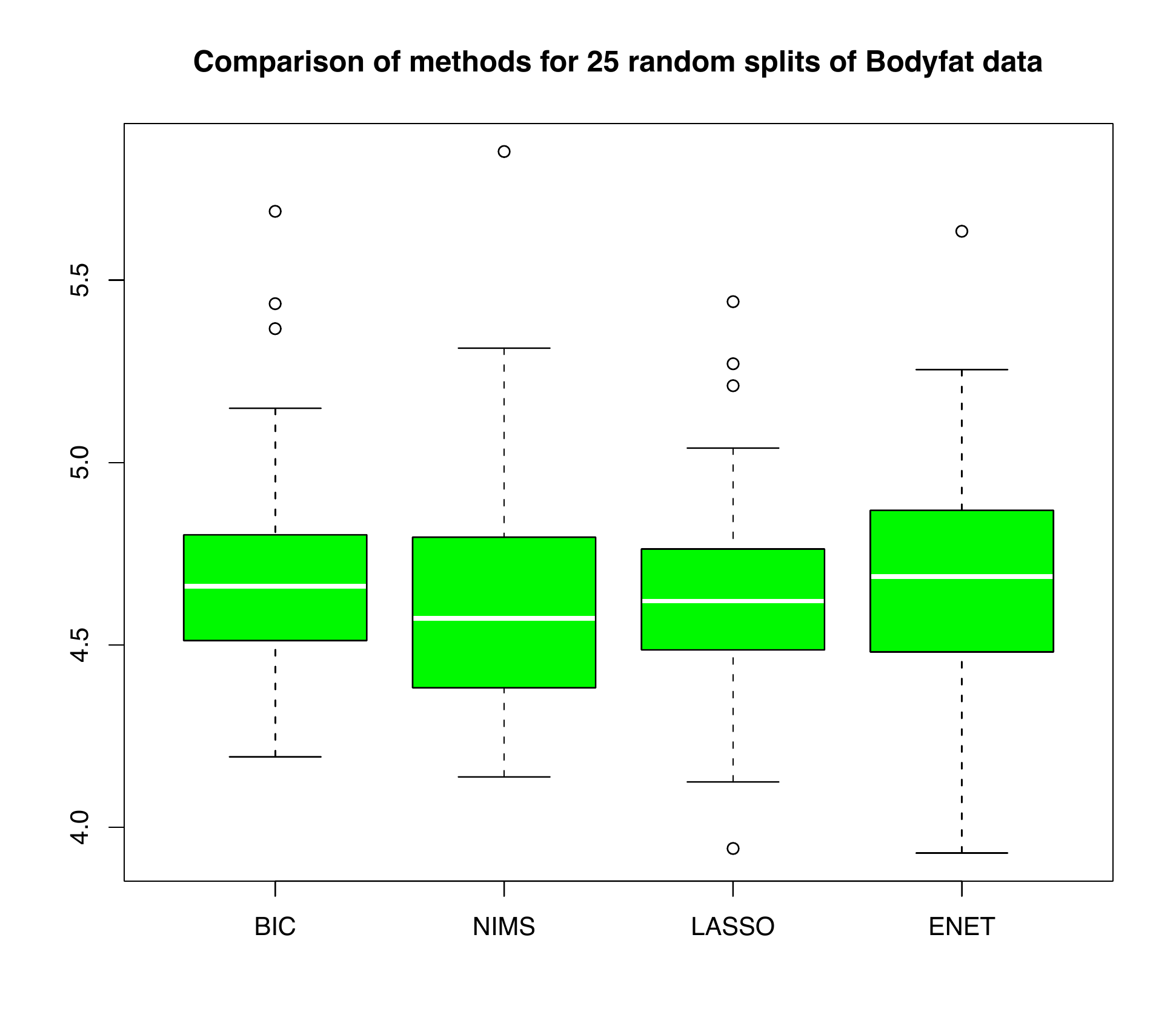}\\
 \includegraphics[width=16cm, height=10cm]{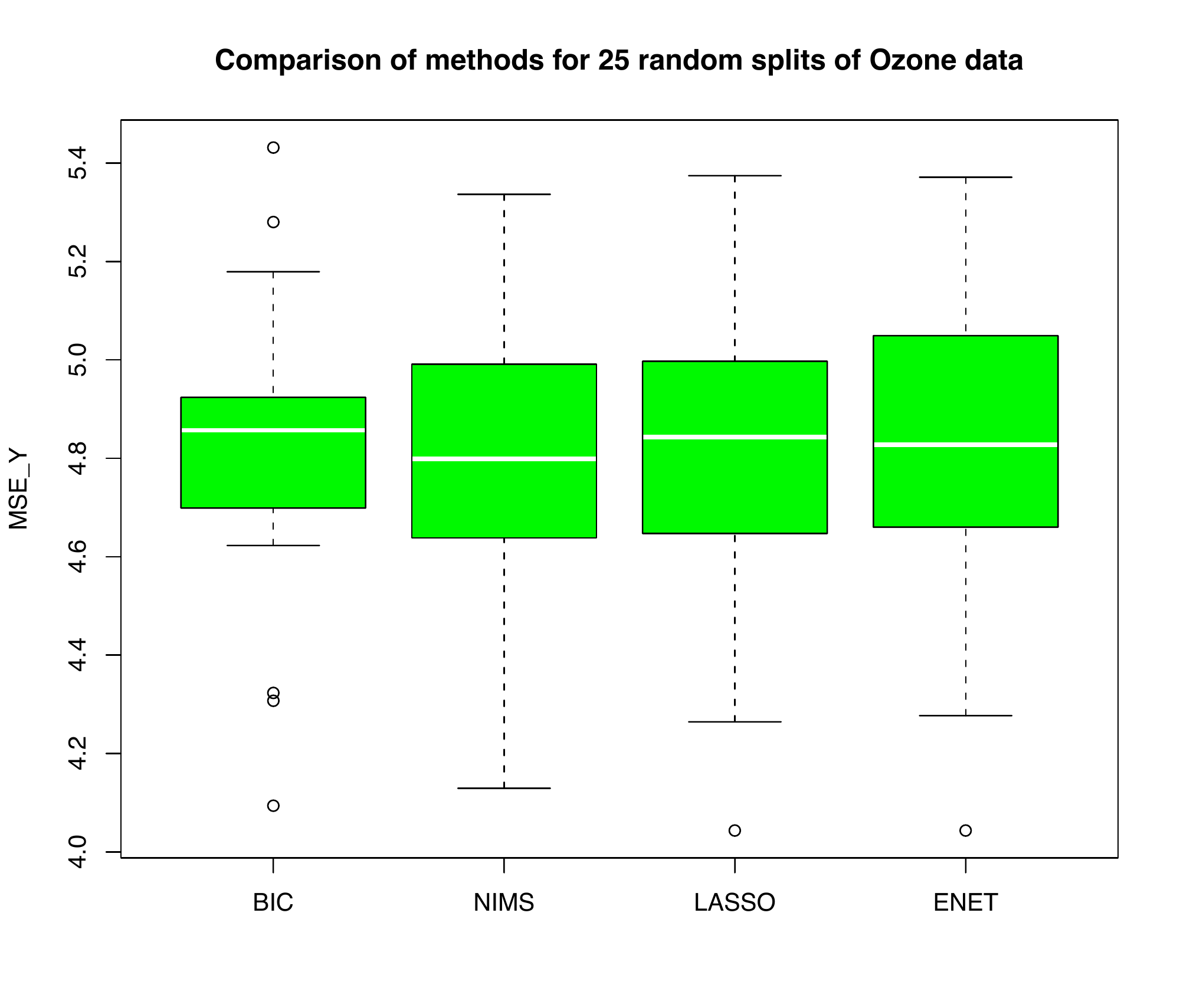}
\end{center}
\caption{Body fat and Ozone datasets: variability of the root mean squared errors over $25$
random splits for BIC, NIMS, LASSO and ENET methods.}\label{boxplot}
\end{figure}

For this dataset, as shown by Table \ref{ozone2}, all Bayesian approaches, as well as AIC and BIC, select about three variables,
while the regularization methods opt for five. The MSE differences  between all procedures are negligible. (This lack of significant differences in the MSEs is
also exhibited through the boxplots of Figure  \ref{boxplot}.)

\section{Conclusion}\label{sec:conc}

\vs In this numerical study, we have compared Bayesian variable selection methods with regularisation methods
in a poorly informative setting. From a variable selection point of view, it appears that the Bayesian methods are more parsimonious
and more relevant than the regularisation methods. From a predictive point of view, there is no significant difference between
both approaches.  Regularisation methods could however be expected to perform better from this latter point of view since they
minimize a cross-validated prediction error. But, owing to model averaging, efficiency, Bayesian methods provide competitive
MSE's.

\newpage

An additional appeal of this study is to single-out and to assess two calibration-free prior models (NIMS and HG-2). 
They both appear as valuable competitors when compared with earlier Bayesian approaches.
However, both methods have a clear drawback (NIMIS is not location invariant and HG-2 excludes the null model).
Nonetheless our series of examples shows that they provide an acceptable objective Bayesian solution for Bayesian
variable selection and regularization in linear models. 

A limitation of this study on our objective Bayesian approach is that we do not consider large
dimensions as in \cite{Bottolo:Richardson:2010}, which require different computational tools to face
the enormous number of potential models. This difficulty is obviously faced by all Bayesian
solutions considered in this paper and is not an issue in terms of the validity of the prior modelling.

\paragraph{Acknowledgments} We are grateful to the Associate Editor and one reviewer for their much valuable comments
and suggestions on a previous version of this paper. They greatly contributed in improving the quality and
presentation of this comparative study.

\begin{table}[htbp]
\begin{center}
\begin{tabular}{|l|c|c|}
  \hline
            & $\mbox{MSE}_{y}$ & Mean \\
            &           & of selected variables \\
  \hline
  AIC       & $4.58 (0.05)$ & $5.56 (0.20)$ \\
  BIC       & $4.60 (0.05)$ & $4.20 (0.18)$ \\
  \hline
  \hline
  BRIC      & $4.51 (0.05)$ & $2.84 (0.15)$ \\
  EB-L      & $4.52 (0.05)$ & $3.00 (0.18)$ \\
  EB-G      & $4.52 (0.05)$ & $3.28 (0.17)$ \\
  ZS-N      & $4.52 (0.05)$ & $2.96 (0.18)$ \\
  ZS-F      & $4.49 (0.05)$ & $4.28 (0.20)$ \\
  OVS       & $4.65 (0.07)$ & $2.96 (0.18)$ \\
  HG-$3$    & $4.54 (0.05)$ & $3.00 (0.18)$ \\
  HG-$4$    & $4.56 (0.05)$ & $3.24 (0.17)$ \\
  \hline
  \hline
  HG-$2$    & $4.50 (0.05)$ & $2.48 (0.14)$ \\
  NIMS       & $4.50 (0.05)$ & $2.44 (0.14)$ \\
  \hline
  \hline
  LASSO     & $4.54 (0.05)$ & $8.17 (0.52)$ \\
  DZ        & $4.51 (0.06)$ & $11.03(0.11)$ \\
  ENET      & $4.54 (0.05)$ & $9.04 (0.56)$ \\
  \hline
\end{tabular}
\caption{\label{body1}
Body fat dataset: Mean of the $\mbox{MSE}_{y}$ and of the selected variables.}
\end{center}
\end{table}

\begin{table}[htbp]
\begin{center}
\begin{tabular}{|l|lllllllllllll|}
  \hline
   Variables & $1$ & $2$    & $3$    & $4$    & $5$    & $6$    & $7$    & $8$    & $9$    & $10$   & $11$   & $12$   & $13$ \\
  \hline
  AIC     & $0.44$ & $0.84$ & $0.16$ & $0.64$ & $0.04$ & $1.00$ & $0.20$ & $0.16$ & $0.08$ & $0.16$ & $0.44$ & $0.80$ & $0.88$ \\
  BIC     & $0.08$ & $0.84$ & $0.08$ & $0.32$ & $0.00$ & $1.00$ & $0.12$ & $0.08$ & $0.04$ & $0.00$ & $0.16$ & $0.28$ & $0.40$ \\
  \hline
  \hline
  BRIC    & $0.08$ & $0.84$ & $0.08$ & $0.32$ & $0.00$ & $1.00$ & $0.12$ & $0.08$ & $0.04$ & $0.00$ & $0.16$ & $0.24$ & $0.40$ \\
  EB-L    & $0.08$ & $0.84$ & $0.08$ & $0.32$ & $0.00$ & $1.00$ & $0.12$ & $0.08$ & $0.04$ & $0.00$ & $0.16$ & $0.28$ & $0.40$ \\
  EB-G    & $0.08$ & $0.88$ & $0.08$ & $0.36$ & $0.00$ & $1.00$ & $0.08$ & $0.08$ & $0.04$ & $0.00$ & $0.20$ & $0.36$ & $0.40$ \\
  ZS-N    & $0.08$ & $0.84$ & $0.08$ & $0.32$ & $0.00$ & $1.00$ & $0.12$ & $0.08$ & $0.04$ & $0.00$ & $0.16$ & $0.24$ & $0.40$ \\
  ZS-F    & $0.20$ & $0.84$ & $0.12$ & $0.40$ & $0.00$ & $1.00$ & $0.12$ & $0.12$ & $0.08$ & $0.04$ & $0.24$ & $0.60$ & $0.68$ \\
  OVS     & $0.12$ & $0.68$ & $0.08$ & $0.16$ & $0.04$ & $1.00$ & $0.08$ & $0.00$ & $0.00$ & $0.00$ & $0.04$ & $0.24$ & $0.52$ \\
  HG-$3$  & $0.08$ & $0.84$ & $0.08$ & $0.32$ & $0.00$ & $1.00$ & $0.12$ & $0.08$ & $0.04$ & $0.00$ & $0.16$ & $0.28$ & $0.40$ \\
  HG-$4$  & $0.08$ & $0.88$ & $0.08$ & $0.32$ & $0.00$ & $1.00$ & $0.08$ & $0.08$ & $0.04$ & $0.00$ & $0.16$ & $0.36$ & $0.40$ \\
  \hline
  \hline
  HG-$2$  & $0.04$ & $0.88$ & $0.00$ & $0.08$ & $0.00$ & $1.00$ & $0.08$ & $0.04$ & $0.00$ & $0.04$ & $0.16$ & $0.28$ & $0.60$ \\
  NIMS     & $0.04$ & $0.88$ & $0.04$ & $0.08$ & $0.00$ & $1.00$ & $0.04$ & $0.08$ & $0.04$ & $0.00$ & $0.04$ & $0.04$ & $0.12$ \\
  \hline
  \hline
  LASSO   & $1.00$    & $0.28$ & $1.00$ & $0.88$ & $0.24$ & $1.00$ & $0.44$ & $0.52$ & $0.28$ & $0.56$ & $0.68$ & $0.84$ & $1.00$ \\
  DZ      & $1.00$    & $0.80$ & $1.00$ & $0.88$ & $0.60$ & $1.00$ & $0.80$ & $0.72$ & $0.40$ & $0.88$ & $0.92$ & $0.88$ & $0.96$ \\
  ENET    & $1.00$    & $0.40$ & $1.00$ & $0.80$ & $0.28$ & $1.00$ & $0.40$ & $0.64$ & $0.44$ & $0.64$ & $0.68$ & $0.84$ & $1.00$ \\
  \hline
\end{tabular}
\caption{\label{body2} Body fat dataset: relative frequencies of selections of the variables over the $25$ random splits+.}
\end{center}
\end{table}

\begin{table}[htbp]
\begin{center}
\begin{tabular}{|l|c|c|}
  \hline
            & $\mbox{MSE}_{y}$ & Mean number \\
            &           & of selected variables \\
  \hline
  AIC       & $4.79 (0.05)$ & $3.52 (0.14)$ \\
  BIC       & $4.77 (0.05)$ & $2.88 (0.07)$ \\
  \hline
  \hline
  BRIC      & $4.78 (0.05)$ & $2.88 (0.07)$ \\
  EB-L      & $4.78 (0.05)$ & $2.88 (0.07)$ \\
  EB-G      & $4.78 (0.05)$ & $2.92 (0.05)$ \\
  ZS-N      & $4.78 (0.05)$ & $2.88 (0.07)$ \\
  ZS-F      & $4.77 (0.05)$ & $3.12 (0.07)$ \\
  OVS       & $4.81 (0.05)$ & $2.88 (0.10)$ \\
  HG-$3$    & $4.78 (0.05)$ & $2.88 (0.07)$ \\
  HG-$4$    & $4.78 (0.05)$ & $2.92 (0.05)$ \\
  \hline
  \hline
  HG-$2$    & $4.80 (0.05)$ & $2.68 (0.10)$ \\
  NIMS      & $4.79 (0.05)$ & $2.68 (0.10)$ \\
  \hline
  \hline
  LASSO     & $4.78 (0.05)$ & $5.24 (0.21)$ \\
  DZ        & $4.80 (0.05)$ & $5.12 (0.13)$ \\
  ENET      & $4.79 (0.05)$ & $5.32 (0.16)$ \\
  \hline
\end{tabular}
\caption{\label{ozone1} Ozone dataset: Mean of the $\mbox{MSE}_{y}$ and of the selected variables.}
\end{center}
\end{table}

\begin{table}[htbp]
\begin{center}
\begin{tabular}{|l|llllllll|}
  \hline
  Variables & $1$ & $2$ & $3$ & $4$ & $5$ & $6$ & $7$ & $8$ \\
  \hline
  AIC     & $0.20$ & $0.12$ & $0.96$  & $1.00$ & $0.56$ & $0.08$ & $0.44$ & $0.16$ \\
  BIC     & $0.04$ & $0.00$ & $0.96$  & $1.00$ & $0.60$ & $0.00$ & $0.36$ & $0.04$ \\
  \hline
  \hline
  BRIC    & $0.04$ & $0.00$ & $0.96$ & $1.00$  & $0.60$ & $0.00$ & $0.40$ & $0.04$ \\
  EB-L    & $0.04$ & $0.00$ & $0.96$ & $1.00$  & $0.60$ & $0.40$ & $0.36$ & $0.04$ \\
  EB-G    & $0.04$ & $0.00$ & $0.96$ & $1.00$  & $0.60$ & $0.00$ & $0.36$ & $0.04$ \\
  ZS-N    & $0.04$ & $0.00$ & $0.96$ & $1.00$  & $0.60$ & $0.00$ & $0.36$ & $0.04$ \\
  ZS-F    & $0.04$ & $0.08$ & $0.92$ & $1.00$  & $0.60$ & $0.00$ & $0.40$ & $0.08$ \\
  OVS     & $0.00$ & $0.00$ & $1.00$ & $0.92$  & $0.00$ & $0.00$ & $0.80$ & $0.08$  \\
  HG-$3$  & $0.04$ & $0.00$ & $0.96$ & $1.00$  & $0.60$ & $0.00$ & $0.36$ & $0.04$ \\
  HG-$4$  & $0.04$ & $0.00$ & $0.96$ & $1.00$  & $0.60$ & $0.00$ & $0.36$ & $0.04$ \\
  \hline
  \hline
  HG-$2$  & $0.04$ & $0.00$ & $0.96$ & $1.00$  & $0.60$ & $0.00$ & $0.32$ & $0.04$ \\
  NIMS    & $0.04$ & $0.00$ & $0.96$ & $1.00$  & $0.60$ & $0.00$ & $0.32$ & $0.04$ \\
  \hline
  \hline
  LASSO   & $0.00$ & $0.00$ & $1.00$ & $1.00$ & $1.00$ & $0.00$ & $1.00$ & $1.00$ \\
  DZ      & $0.00$ & $0.00$ & $1.00$ & $1.00$ & $1.00$ & $0.00$ & $1.00$ & $1.00$ \\
  ENET    & $0.00$ & $0.00$ & $1.00$ & $1.00$ & $1.00$ & $0.00$ & $1.00$ & $1.00$ \\
  \hline
\end{tabular}
\caption{\label{ozone2}
Ozone dataset: relative frequencies of selections of the variables over the $25$ random splits.}
\end{center}
\end{table}

\end{document}